\documentclass[pre,aps,showpacs,supergroupedaddress,epsfig,amsmath,amssymb,eqsecnum]{revtex4}
\usepackage{epsfig,amsmath,amssymb,bm,epsf,graphics,psfrag,verbatim,subfigure,framed}
\usepackage[usenames,dvipsnames]{color}
\usepackage{bbm}
\usepackage{mathrsfs}
\usepackage{amsfonts}
\usepackage{color}
\usepackage{pbox}
\def\tv{\widetilde{v}}

\def\tv{\widetilde{v}}

\newcommand{\be}{\begin{equation}}
\newcommand{\ee}{\end{equation}}
\newcommand{\ba}{\begin{eqnarray}}
\newcommand{\ea}{\end{eqnarray}}
\newcommand{\bw}{\begin{widetext}}
\newcommand{\ew}{\end{widetext}}

\newcommand{\rv}{{\mathbf{r}}}
\newcommand{\dv}{{\mathbf{d}}}
\newcommand{\xv}{{\mathbf{x}}}

\newcommand{\kv}{{\bm{k}}}
\newcommand{\pv}{{\mathbf{p}}}

\newcommand{\cv}{{\mathbf{c}}}
\newcommand{\lv}{{\mathbf{\lambda}}}

\newcommand{\Rv}{{\mathbf{R}}}

\begin{document}

\title{Van der Waals interactions between polymers with sequence-specific polarizabilities: Stiff polymers and Gaussian coils}
\author{Bing-Sui Lu$^1$}
\email{bing-sui.lu@fmf.uni-lj.si}
\author{Ali Naji$^2$}
\author{Rudolf Podgornik$^1$}
\affiliation{$^{1}$Department of Theoretical Physics, J. Stefan Institute, 1000 Ljubljana, Slovenia.\\
 $^{2}$School of Physics, Institute for Research in Fundamental Sciences (IPM), P.O. Box 19395-5531, Tehran, Iran
}

\date{\today}


\begin{abstract}
We consider the van der Waals interaction between a pair of polymers with quenched heterogeneous sequences of local polarizabilities along their backbones, and study the effective pairwise interaction energy for both stiff polymers and flexible Gaussian coils. In particular, we focus on the cases where the pair of polarizability sequences are (i)~distinct and (ii)~identical.  
We find that the pairwise interaction energies of distinct and identical Gaussian coils are both isotropic and exhibit the same decay behavior for separations larger than their gyration radius, in contradistinction to the orientationally anisotropic interaction energies of distinct and identical stiff polymers. For both Gaussian coils and stiff polymers, the attractive interaction between identical polymers is enhanced if the polarizability sequence is more heterogeneous. 
\end{abstract}

\maketitle 

\section{Introduction}

Van der Waals (vdw) forces\cite{French,Dalvit,Bordag,Parsegian} are long ranged forces which act between all kinds of bodies including electrically neutral ones, and are therefore prevalent in nature. These forces lead to a variety of phenomena including flocculation in colloid systems,\cite{verwey_overbeek} and are also responsible for the ability of geckos to stick to walls.\cite{gecko} Van der Waals forces furthermore appear to be active in the first stages of planet formation, in causing planetisimals to cohere together when the gravitational force is still too weak to bring about cohesion.\cite{cuzzi} In industry and water treatment plants, vdw forces have also been utilized to treat waste water. Waste water contains impurities which are charged, and coagulants are added to make them neutral. These neutral impurities which are colloidal then attract each other by vdw forces and settle down as floc which can then be easily disposed of.\cite{wastewater} 

More relevantly for evolutionary biology and the physics of self-assembly, vdw forces also contribute to the process of recognition between molecules. Molecular recognition is the process whereby a (bio-)molecule, say a protein, 
recognizes a specific structural feature on another macromolecule that it interacts strongly with.\cite{neidle,cherstvy} 
This includes the recognition between proteins and DNA molecules that are so essential for biological functions. A striking feature of biological molecular recognition is its high target sequence specificity,\cite{lukatsky1,lukatsky2} for example, certain protein molecules such as the lac repressor are able to bind to a specific sequence out of six million possible ones on the DNA of an \emph{E. coli} bacterium. On the other hand, in molecular electronics and DNA computation,\cite{adleman} there has been increasing interest in the possibility of self-assembling DNA-based circuit boards, the self-assembly being facilitated by the molecular recognition between pairs of DNA molecules.\cite{seeman} An investigation of vdW interactions between polymers would thus be of relevance to better understanding the process of molecular recognition.
 
Our paper is structured as follows. We first look (heuristically) at how vdW interactions can arise between neutral but polarizable atoms. We then consider physical aspects of the dsDNA molecule, which motivates our study of vdW interactions between stiff polarizable polymers. Next, we explain the formalism we use to describe such vdW interactions, focussing on the features that can aid the mechanism of molecular recognition.\cite{LNP} Towards the end we will also consider the case of flexible polarizable Gaussian coils (which can describe globular proteins such as lysozyme and RNA polymerase), and how their vdW interaction differs from that between stiff polymers.

\subsection{Van der Waals interactions: Heuristics}
First, let us consider heuristically how vdW interactions can arise. Let us consider a neutral atom, whose proton charge balances the charge of its electron cloud. 
At finite temperature, the electron cloud undergoes thermally driven distortions, and at any one instant the center of the electron cloud will be displaced relative to the position of the proton. It thus acquires an induced dipole moment. A second atom in the vicinity of the first atom experiences an electric field generated by the induced dipole of the first atom, which polarizes the electron cloud of the second atom, giving rise to another induced dipole. Although each induced dipole vanishes under time averaging, the product of two induced dipoles does not vanish, and this gives rise to a non-zero polarization-type or van der Waals interaction between the atoms. 

\subsection{Structure of dsDNA molecules}
Next, let us look at some structural features of dsDNA molecules that are relevant to molecular recognition.\cite{alberts} The molecules have persistence lengths of around 100 base pairs (about 50 nm), which means that short dsDNA molecules (e.g., 10 to 20 base pairs long) essentially behave as stiff rodlike molecules. Each dsDNA molecule, having a length that is much greater than its radius, is correspondingly much more polarizable along its backbone than in the directions transverse to it. Additionally, each dsDNA molecule comes equipped with a sequence of different nucleotide base pairs, different base pairs having different polarizabilities depending on the type of base pair and the identity of its neighbours.\cite{schimelman} In solution the highly charged dsDNA molecule also attracts counterions to the backbone, which further modifies the local polarizability along the backbone. All this implies that the associated sequence of polarizabilities is heterogeneous along the backbone of the dsDNA molecule. Significantly, as the base pair sequence is structurally built into the backbone of the molecule, the corresponding sequence of polarizabilities is also thereby \emph{quenched} (i.e., the polarizabilities are effectively fixed, albeit random, on experimental time-scales). In general there will also be an annealed (i.e., thermally fluctuating) contribution to the local polarizability arising from the condensed counterions, but for simplicity, we only consider the situation in which the polarizabilities are entirely quenched. 

The base pairs are hydrophobic, which means that they carry induced dipoles and do not carry permanent dipoles. On the other hand, the sugar phosphate backbone of the dsDNA molecule is highly negatively charged with a linear charge density of around 6 elementary charges per nm. In solution, this induces a cloud of counterions to condense on the backbone bringing down the net charge of the dsDNA to an effective charge of around 1.4 elementary charges per nm in the case of monovalent counterions.\cite{manning,boroudjerdi} The electrostatic potential due to this effective charge is further screened through the Debye screening mechanism. Associated with the latter effect is the Debye screening length, defined by $\lambda_{{\rm D}} = (\varepsilon \varepsilon_0 k_{{\rm B}}T / \sum_{j} c_j q_j^2)$, where $\varepsilon$ is the relative permittivity of the solvent medium (for water, $\varepsilon\approx 80$), $\varepsilon_0$ is the vacuum permittivity, $k_{{\rm B}}$ is Boltzmann's constant, $T$ is the temperature, $c_j$ is the concentration of ion species $j$, and $q_j$ the corresponding charge valence of the ion species. For example, at room temperature ($T=298 \, \rm{K}$) and for a salt solution containing only NaCl of concentration $1 \, \rm{M}$
%$1 \text{mol} \, \text{dm}^{-3}$
the corresponding Debye length is given by $\lambda_{{\rm D}} \approx 0.3 \, \text{nm}$. At separation distances greater than $\lambda_{{\rm D}}$, electrostatic interactions are much less important than vdW interactions, and the molecule is effectively neutral. We consider inter-molecule separations of order $1$ to tens of nm, which means that we can essentially disregard electrostatic interactions and focus only on vdW interactions. 

\subsection{Objective and strategy}
\label{sec:objective}
In what follows, we are going to address the following questions: 
(i)~How does the vdW interaction between two polymers depend on the correlatedness of their polarizability sequences? Specifically, is the vdW interaction statistically stronger between a pair of polymers with identical polarizability sequences than between a pair of polymers with distinct sequences? 
(ii)~What effect does a more pronounced polarizability heterogeneity have on vdW attraction?
(iii)~How does the vdW interaction depend on the relative orientation of two polymers?  
We will see that the vdW interaction between a pair of stiff polarizable polymers is characterized by the interplay between polarizability sequence specificity, heterogeneity, and orientational anisotropy, all of which contribute to the mechanism of molecular recognition. 
\begin{figure}
\begin{center}
		\includegraphics[width=0.36\textwidth]{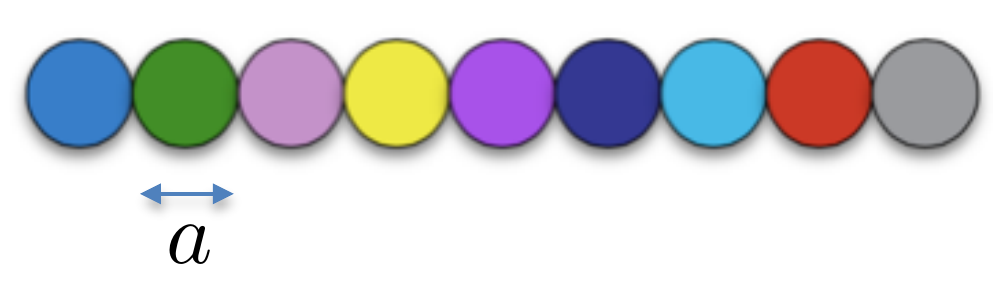}
		\end{center}
	\caption{``Shish-kebab" model of a dsDNA molecule as a stiff polymer consisting of monomers (i.e., nucleotide base pairs, represented by circles) of equal size $a$. The differences in shade reflect differences in polarizability.} 
\label{fig:shishkebab}
\end{figure}

For the rest of the paper, we adopt the following strategy. Firstly, we approximate a dsDNA molecule by the shish-kebab model.\cite{doi-edwards} In this model, the polymer is a stiff rod, and each monomer (i.e., a nucleotide base pair) is assumed to have the same size $a$. The monomers are represented in Fig.~\ref{fig:shishkebab} by circles with different shades, corresponding to differences in polarizability. Each monomer can have a different polarizability value.  
As dsDNA molecules are non-zwitterionic, we only consider interactions between induced dipoles.
We have also approximated the polarizability by its static value, an approximation that works well at high temperature. On the other hand, we consider only non-retarded interactions, i.e., corrections due to the finite speed of light can be neglected, and this is a good approximation at inter-molecule separations smaller than the retardation lengthscale $\hbar c/k_{{\rm B}}T$ (obtained, e.g., by comparing the London dispersion formula and Casimir-Polder formula for a pair of atoms). Thus we see that the non-retarded approximation is only valid for temperatures that are not too high. For biological systems, $T\approx 300 \,{\rm K}$, which corresponds to a retardation lengthscale of $7 \,\mu {\rm m}$. The range of separations that we are interested in probing (one to tens of nm) thus falls safely within the non-retarded regime.
It is well known that water has structure (water has also been regarded as a network of hydrogen bonds). To make the problem analytically tractable, we approximate the aqueous medium as a homogeneous dielectric of $\varepsilon \approx 80$.

\section{Hamiltonian For Induced Dipoles}
\subsection{Dipole electrostatics}
To describe the vdW interaction between polarizable polymers, we proceed by constructing a Hamiltonian for the induced dipoles. This requires us to first define the dipole charge density. In a neutral molecule, if the charge of the proton is $q$ and the separation vector between the proton and the center of the electron cloud is $\xv$, the dipole moment is given by $\mathbf{p} = q \, \xv$. The corresponding dipole charge density is given by 
$\rho(\rv) = -q\delta(\rv-\Rv+\frac{\xv}{2}) +q\delta(\rv-\Rv-\frac{\xv}{2})$, where $\Rv$ denotes the position vector of the centre of the dipole. For small $d$, we can do a Taylor expansion. This yields $\rho(\rv) \approx -\pv\cdot\nabla\delta(\rv-\Rv)$. 
For $N$ polymers each with $M$ monomers, the dipole charge density is given by 
\be
\rho(\rv) = -\sum_{i=1}^{N}\sum_{m=1}^{M} \pv^{(i)}(m)\cdot\nabla\delta(\rv-\Rv^{(i)}(m)), 
\ee
where now $\pv^{(i)}$ ($\Rv^{(i)}(m)$) refers to the induced dipole moment (position vector) of monomer $m$ on polymer $i$. 
Having defined the dipole charge density, we can write down the standard expression for the potential at position $\rv$, viz.,
\be
\Phi(\rv) = \frac{1}{4\pi\varepsilon\varepsilon_0} \int d^3\rv \frac{\rho(\rv')}{|\rv-\rv'|}. 
\ee 
The electrostatic energy due to dipoles is given by 
\ba
H_{p,N} &=& \frac{1}{2}\int d^3\rv \, \rho(\rv) \Phi(\rv)
\nonumber\\
&=&
\frac{1}{8\pi\varepsilon\varepsilon_0} 
\sum_{i \neq j}^{N} \sum_{m\neq n}^{M} 
\left\{ 
\frac{\pv^{(i)}(m)\cdot\pv^{(j)}(n)}{(R_{ij}(m,n))^3} \right.
\nonumber\\
&& 
\left.-
\frac{3(\pv^{(i)}(m)\cdot\Rv_{ij}(m,n))(\pv^{(j)}(n)\cdot\Rv_{ij}(m,n))}{(R_{ij}(m,n))^5}
\right\}.
\ea
Here $\Rv_{ij}(m,n) \equiv \Rv^{(j)}(n) - \Rv^{(i)}(m)$, the factor of $1/2$ is to prevent double counting in the sum over pairwise interactions of dipoles. By plugging in our formula for the dipole charge density, we find that the interaction potential between two dipoles is equal to two gradients acting on the Coulomb potential between two monopolar charges. 

\subsection{Dipole ``elastics"}
In addition to the foregoing \emph{electrostatic} term describing the interactions between dipoles, we also have an \emph{elastic} term describing the energetic cost of distorting the electron cloud of each molecule. If we have an external field $\mathbf{E}$, the field will induce a dipole moment $\pv$, and these quantities are related via the equation $\pv=\mathbf{\alpha} \cdot \mathbf{E}$, where $\mathbf{\alpha}$ is the polarizability \emph{tensor}. A larger value of $\mathbf{\alpha}$ means that the molecule is more polarizable. Rewriting $\mathbf{E} = \mathbf{\alpha}^{-1} \cdot \pv$, we see that this relation is reminiscent of Hooke's law, and thus we can regard $\mathbf{\alpha}^{-1}$ as a ``spring constant" and $\pv$ as an ``extension." As the polymer is much more polarizable along its backbone than in directions perpendicular to it, we can approximate $\mathbf{E} \approx \alpha^{-1} \pv$, where $\alpha$ is now the polarizability along the backbone.  By analogy with Hookean elasticity, we can write the ``dipole elastic distortion energy" for a monomer, viz., 
\be
H_{e} = \frac{1}{2} \alpha^{-1} p^2. 
\ee
For $N$ polymers each with $M$ monomers, the dipole elastic distortion energy is given by 
\be
H_{e,N} = \sum_{i=1}^{N}\sum_{m=1}^{M} \frac{(p^{(i)}(m))^2}{\alpha^{(i)}(m)}.
\ee
The full Hamiltonian for $N$ polymers is then given by 
\be
H_{N} = H_{p,N} + H_{e,N}. 
\ee
As biological systems are typically at room temperature, there are many different thermal configurations of the induced dipoles, each characterized by a different Boltzmann weight. The effective interaction energy between a pair of \emph{rigid} (i.e., conformationally non-fluctuating) polymers comes about by coarse-graining over their respective induced dipole degrees of freedom. 
As the dipoles are much more easily induced along the polymer backbone, we can equate the direction of the dipole to the orientation of the polymer, given by the tangent vector $\tv^{(i)}$. For \emph{stiff} polymers, the direction of each dipole is constrained by the orientation of the polymer, and thus for a given polymer conformation only the magnitudes of the dipoles undergo thermal fluctuations. Thus we can write the partition function for a pair of rigid stiff polymers as 
\be
Z(N, \{ \Rv^{(i)} \}) = \prod_{i,m} \int dp^{(i)}(m) e^{-\beta H_N}. 
\ee
It is very difficult to evaluate the partition function exactly, owing to the large number of interacting dipoles. One method of making the problem analytically tractable is the Hubbard-Stratonovich transformation method~(see, e.g., Ref.~\cite{zinn-justin}), where we introduce a local auxiliary field $\varphi$ to decouple the dipoles. The dipoles can then be integrated out, and what remains is a local theory for $\varphi$,\cite{LNP} viz.,
\be
Z(N, \{ \Rv^{(i)} \}) \propto \! \int \! \mathcal{D}\varphi\, \, e^{-\frac{1}{2} \beta\int\! d\rv \,\varepsilon\varepsilon_0(\nabla\varphi)^2}
 e^{-\frac{1}{2}\beta\sum\limits_{i,m} \alpha^{(i)}(m) \big( \tv^{(i)}(m) \cdot \nabla\varphi(\Rv^{(i)}(m)) \big)^{2}}.
\label{eq:part_function:2}
\ee
The resulting Hamiltonian is quadratic in the gradient of $\varphi$. We can interpret $\varphi$ as the local field fluctuation generated by the presence of the induced dipoles. We also see that the dipoles effectively contribute to the dielectric permittivity of the polymer: away from the polymer, the dielectric permittivity is $\varepsilon\varepsilon_0$, and at the position of the polymer, there is a correction proportional to the dipole polarizability.  

\section{Effective Interaction Energy}
The calculation of the partition function can be simplified if the polarizability is small. We can then expand the exponent in powers of $\alpha$ up to second order, the second order being required to see the interaction between the pair of polymers. This is also formally equivalent to making a Hamaker summation approximation. We find\cite{LNP} 
\be
Z(N, \{ \Rv^{(i)} \}) \approx e^{\beta(F_{{\rm self}} + F_{{\rm int}})},
\label{eq:hamakera}
\ee
where $F_{{\rm self}}$ is the sum of the free energies of individual polymers and $F_{{\rm int}}$ is the effective interaction energy between pairs of polymers. For a pair of polymers, we find
\ba
F_{{\rm int}} &=& -\frac{k_{{\rm B}}T}{32\pi^2(\varepsilon\varepsilon_0)^2 a^2} 
\int ds_1 \int ds_2 \, \alpha^{(1)}(s_1) \alpha^{(2)}(s_2) 
\nonumber\\
&&
\times 
\left\{ 
\frac{\tv^{(1)}(s_1)\cdot\tv^{(2)}(s_2)}{R_{12}^3} - \frac{3(\tv^{(1)}(s_1)\cdot\Rv_{12})(\tv^{(2)}(s_2)\cdot\Rv_{12})}{R_{12}^5}
\right\}^2,
\label{eq:fint_pair}
\ea
where $\Rv_{12} \equiv \rv - s_2 \tv^{(2)} + s_1 \tv^{(1)}$, $\tv^{(1)}$ and $\tv^{(2)}$ are the tangent vectors of the first and second polymers, $\rv$ is the separation between their centers of mass, and $s_1$ and $s_2$ are the arc lengths of the first and second polymers. In deriving the above result, we have taken the continuum limit, where the sum over segments becomes an integral over the arc length. We note that inside the integrand the interaction decays as $R^{-6}$, which is the usual decay law associated with non-retarded dispersion interactions.

In the effective interaction energy, we also need to average over the polarizabilities. We assume that the local polarizabilities follow a Gaussian distribution, and the local polarizability is composed of a mean and a quenched fluctuation, i.e., $\alpha^{(i)}(s) = \alpha_0 + \delta \alpha^{(i)}(s)$. The mean $\alpha_0$ and variance $g^2$ specified by $\langle\!\langle  \alpha^{(i)}(s) \rangle\!\rangle = \alpha_0$ and $\langle\!\langle \delta\alpha^{(i)}(s) \delta\alpha^{(j)}(s') \rangle\!\rangle = g^2 a \delta(s-s') \delta_{ij}$, where $\langle\!\langle  \ldots \rangle\!\rangle$ denotes a \emph{sequence average} over the statistics of the polarizability. From such a statistics, we can deduce the following two limiting cases. For a pair of polymers with \emph{distinct} sequences (i.e., $\alpha^{(i)}(s) \neq \alpha^{(j)}(s)$), this implies that $\langle\!\langle \alpha^{(i)}(s) \alpha^{(j)}(s') \rangle\!\rangle = \alpha_0^2$, whereas for a pair of polymers with \emph{identical} sequences (i.e., $\alpha^{(i)}(s) = \alpha^{(j)}(s) \equiv \alpha(s)$), the polarizability correlator receives an additional contribution from the variance, viz., $\langle\!\langle \alpha^{(i)}(s) \alpha^{(j)}(s') \rangle\!\rangle = \alpha_0^2 + g^2 a \delta(s-s')$. 

In what follows, we first give an overview of the main results of Ref.~\cite{LNP}, for the interaction behavior of two stiff polymers, in the two limiting cases where the polymers have \emph{distinct} and \emph{identical} polarizability sequences. We then consider the interaction behavior of two flexible polarizable Gaussian coils. 

\section{Results}
\subsection{Two stiff polymers}
Let us first consider stiff polymers with \emph{distinct} sequences. In the near-field regime (by which we mean their separation distance is much smaller than the length of each polymer), we can approximate each polymer by an infinitely long polymer. From Eq.~(\ref{eq:fint_pair}), we have found that the sequence averaged interaction energy is given by\cite{LNP} 
\be
\langle\!\langle F_{{\rm int}} \rangle\!\rangle \approx - \frac{M^2\alpha_0^2 k_{{\rm B}}T}{64\pi(\varepsilon\varepsilon_0)^2\ell^2} \frac{z^2}{\sqrt{1-z^2}}\frac{1}{|\Rv_{12}^\ast|^4},
\ee
where $z = \tv^{(1)}\cdot\tv^{(2)}$, $\ell$ is the length of each polymer, and $\Rv_{12}^\ast$ is the shortest length vector that connects the two polymers. 
In the far-field regime (i.e., where the separation distance is much greater than the length of each polymer), we have found that\cite{LNP}  
\be
\langle\!\langle F_{{\rm int}} \rangle\!\rangle \approx - \frac{(z-3y_1y_2)^2M^2\alpha_0^2k_{{\rm B}}T}{32\pi^2(\varepsilon\varepsilon_0)^2r^6},
\ee
where $y_1 = \tv^{(1)}\cdot \rv/r$ and $y_2 = \tv^{(2)}\cdot \rv/r$. Note that in both the near- and far-field regimes, $\langle\!\langle F_{{\rm int}} \rangle\!\rangle$ is invariant with respect to inversion of each polymer about its center, as there is no correlation of polarizability fluctuations for the two distinct polymers, and thus each polymer sees a uniform (sequence averaged) polarizability on the other polymer.

Next, we consider the sequence averaged effective interaction energy for a pair of \emph{identical} sequences. As before, we first look at the behavior in the near-field regime. The sequence averaged energy is now given by 
\be
\langle\!\langle F_{{\rm int}} \rangle\!\rangle \approx F_0 + \delta F,
\ee
where 
\ba
F_0 &=& - \frac{M^2\alpha_0^2 k_{{\rm B}}T}{64\pi(\varepsilon\varepsilon_0)^2\ell^2} \frac{z^2}{\sqrt{1-z^2}}\frac{1}{|\Rv_{12}^\ast|^4}; 
\\
\delta F &=&
- 
\frac{3k_{{\rm B}}TMg^2\chi(y_1,y_2,z)}{16384\sqrt{2}\pi(\varepsilon\varepsilon_0)^2 \ell r^5}.
\ea
We find two contributions: the first ($F_0$) is the contribution coming from the mean polarizability, and is identical with the sequence averaged interaction energy of two polymers with distinct sequences, whereas the second contribution ($\delta F$) is new, and comes from the correlation of quenched polarizability fluctuations. Note that $\delta F$ decays as $r^{-5}$.  
In the result above, $\chi$ describes the orientational anistropy of the vdW interaction, and is given by 
\ba
\chi(y_1,y_2,z) &=&
\frac{1}{\sqrt{1-z} \left( 1 - \gamma \right)^{5/2}}
\bigg\{
9+14z+41z^2 
\nonumber\\
&& 
- \frac{5(3+10z+3z^2)(y_1+y_2)^2}{1 - \gamma}
+\frac{105(y_1+y_2)^4}{4\left( 1 - \gamma \right)^2}
\bigg\}.
\label{eq:chi_orientation}
\ea
Here, $\gamma \equiv \frac{(y_1-y_2)^2}{2(1-z)}$, and distinguishes between polymers that are aligned and those that are anti-aligned. 
We can distinguish between a state of alignment and a state of anti-alignment, because now the sequences are perfectly correlated, and to flip one sequence you have to flip the other too to get the same energy.

In the far-field regime, we find that the sequence averaged effective interaction energy is given by 
\be
\langle\!\langle F_{{\rm int}} \rangle\!\rangle \approx -\frac{k_{\mathrm{B}}T(M^2 \alpha_0^2+ M g^2)(z - 3 \, y_1y_2)^2}{32 \pi^2(\varepsilon\varepsilon_0)^2 r^6} + O((\ell/r)^2),   
\label{eq:deltaF_far}
\ee
where we see that the only effect of the polarizability correlation is to renormalize the overall prefactor. 

\subsection{Two flexible Gaussian coils}

We now address the question of how the interaction behavior is modified when the pair of rigid stiff polymers are replaced by a pair of flexible Gaussian chains. For such chains, the local tangent vector $\tv^{(i)}(s)$ of the coil and the polymer coordinate $\Rv^{(i)}(s)$ are independent of each other, in contradistinction to the case of stiff polymers. Let us write $F_{{\rm int}}$ for the effective interaction energy for a given chain conformation. Recognizing the complexity of a full-fledged description of interaction between \emph{given conformations} of Gaussian chains, the problem can be simplified by going to a coarse-grained description of interaction between Gaussian \emph{coils}, where the thermal averaging over polymer coordinates and local tangent vectors has already been carried out on the partition function $Z$. This leads to 
\ba
\langle \overline{Z(N, \{ \Rv^{(i)}, \tv^{(i)} \})} \rangle_{\{\Rv^{(i)}\}} &\equiv& \prod_{i,s}\int \! d\Rv^{(i)}(s) \! \int \! d\tv^{(i)}(s) Z[\{ \Rv^{(i)}(s), \tv^{(i)}(s) \}]
\nonumber\\
&&\propto 1-\beta (\langle \overline{F_{{\rm self}}}\rangle_{\{\Rv^{(i)}\}} + \langle \overline{F_{{\rm int}}}\rangle_{\{\Rv^{(i)}\}})
\nonumber\\
&&\approx e^{-\beta (\langle \overline{F_{{\rm self}}}\rangle_{\{\Rv^{(i)}\}} + \langle \overline{F_{{\rm int}}}\rangle_{\{\Rv^{(i)}\}})}.
\ea 
In the second step we made use of the second order approximation as in Eq.~(\ref{eq:hamakera}), and in the third step we re-exponentiated the terms. The overhead bar denotes averaging with respect to the local tangent vectors, and $ \langle \ldots \rangle_{\{\Rv^{(i)}\}}$ denotes averaging with respect to polymer coordinates $\{\Rv^{(i)}\}$. The sequence average can also be performed on $\langle \overline{F_{{\rm int}}}\rangle_{\{\Rv^{(i)}\}}$. 

In the high temperature regime, we can make a further simplification by making use of the rotating dipole approximation,\cite{ksm} which consists of (isotropically) averaging $F_{{\rm int}}$ over all orientations in space. This approximation is valid when the Gaussian coils are spherical, which is the case when the centers of mass of the coils are sufficiently far apart, i.e., by a distance larger than the effective radius of gyration of the coils (defined in Eq.~(\ref{eq:eff_gyration_r})). 

To perform the orientation average, we make use of the following relations valid for Gaussian chains:
\ba
&&\overline{t_a^{(1)}(s) \, t_b^{(2)}(s')} = 0;
\\
&&\overline{t_a^{(1)}(s) \, t_b^{(1)}(s)} = \overline{t_a^{(2)}(s) \, t_b^{(2)}(s)} = \delta_{ab},
\ea
where $a, b = 1,2,3$ are Cartesian indices. 
After averaging $F_{{\rm int}}$ [cf. Eq.~(\ref{eq:fint_pair})] over the orientations, we obtain 
\be
\overline{F_{{\rm int}}} = -\frac{3 k_{{\rm B}} T}{16\pi^2(\varepsilon\varepsilon_0)^2} \int \! \frac{ds}{a}  \! \int  \! \frac{ds'}{a} 
\frac{{\alpha}^{(1)}(s) \, {\alpha}^{(2)}(s')}{(R_{12}(s,s'))^6}, 
\label{eq:Fint_coil}
\ee
where $\Rv_{12}(s,s') \equiv \Rv^{(2)}(s') - \Rv^{(1)}(s)$. 
We still need to perform the average over fluctuating polymer coordinates as well as the sequence average over fluctuating polarizabilities. 
Parallel to our analysis for the stiff polymers, we can study the interaction behavior of a pair of Gaussian coils when (i)~their separation distance is much larger than $R_g$ (the effective radius of gyration of the coils, which is of nanometer order, cf. Eq.~(\ref{eq:eff_gyration_r})) but still smaller than the retardation lengthscale (which is of micron order at $T=300 \,{\rm K}$), and (ii)~when the coils are near each other at separations $\gtrsim R_g$, in order that the Gaussian coil conformations still follow an isotropic Gaussian distribution. For each regime, we consider the following two cases: (i)~distinct and (ii)~identical polymers. 

\subsubsection{Coils far apart}

We consider the regime in which two coils are separated by a distance much greater than their effective radius of gyration. Let us first study the case of \emph{distinct} polymers, for which there is no correlation of polarizabilities between the coils. 
The sequence average of the interaction free energy in Eq.~(\ref{eq:Fint_coil}) thus yields
\be
\langle\!\langle \overline{F_{{\rm int}}} \rangle\!\rangle = -\frac{3 k_{{\rm B}} T\alpha_0^2}{16\pi^2(\varepsilon\varepsilon_0)^2} \int \! \frac{ds}{a}  \! \int  \! \frac{ds'}{a} 
\frac{1}{(R_{12}(s,s'))^6}. 
\label{eq:F0_coil}
\ee
The centers of mass of the first and second Gaussian coils are given by
$\Rv_G^{(1)} \equiv \int \! \frac{ds}{\ell} \Rv^{(1)}(s)$ and $\Rv_G^{(2)} \equiv \int \! \frac{ds}{\ell} \Rv^{(2)}(s)$.  
Let us define the fluctuations $\delta \Rv^{(i)}(s) = \Rv^{(i)}(s) - \Rv_G^{(i)}$ and $\delta\Rv(s,s') \equiv \Rv_{12}(s,s') - (\Rv_G^{(2)} - \Rv_G^{(1)}) = \delta\Rv^{(2)}(s') - \delta\Rv^{(1)}(s)$. 
In the far-field regime, $\delta R \ll d$, where $\dv \equiv \Rv_G^{(2)} - \Rv_G^{(1)}$ is the separation between the centers of mass of the Gaussian coils, and we can perform a Taylor expansion (to quadratic order in $\delta\rv$): 
\be
\langle\!\langle \overline{F_{{\rm int}}} \rangle\!\rangle \approx 
-\frac{3 k_{{\rm B}} T\alpha_0^2}{16\pi^2(\varepsilon\varepsilon_0)^2 d^6} 
\int \! \frac{ds}{a}  \! 
\int  \! \frac{ds'}{a} 
\left\{ 
1 - \frac{3 |\delta\Rv|^2}{d^2} 
+ \frac{24 (\dv\cdot\delta\Rv)^2}{d^4} 
\right\}.
\ee
To make further progress, we assume the thermal fluctuations $\delta\Rv^{(1)}$ and $\delta\Rv^{(2)}$ are Gaussian distributed, and average $\langle\!\langle \overline{F_{{\rm int}}} \rangle\!\rangle$ over these fluctuations. 
The radii of gyration of the individual coils are given by $R_g^{(1)}$ and $R_g^{(2)}$, which are defined by $(R_g^{(i)})^2 \equiv \int \! \frac{ds}{\ell} \langle |\delta\Rv^{(i)}(s)|^2 \rangle_{R_i}$. 
For isotropic coils, we have 
\begin{subequations}
\ba
&&\int \frac{ds}{\ell} \langle \delta R_a^{(1)}(s) \, \delta R_b^{(1)}(s) \rangle_{R_1} = \frac{1}{3}\delta_{ab}(R_g^{(1)})^2;
\\
&&\int \frac{ds}{\ell} \langle \delta R_a^{(2)}(s') \, \delta R_b^{(2)}(s') \rangle_{R_2} = \frac{1}{3}\delta_{ab}(R_g^{(2)})^2;
\\
&&\int \frac{ds}{\ell} \langle \delta \Rv^{(1)}(s) \rangle_{R_1} = \langle \delta \Rv^{(2)}(s') \rangle_{R_2} = 0.
\ea
\label{eq:coil_iso_stats}
\end{subequations}
Performing the average over polymer coordinates, the interaction free energy becomes
\be
\langle \langle\!\langle \overline{F_{{\rm int}}} \rangle\!\rangle \rangle_{R_1, R_2} =
-\frac{3 k_{{\rm B}} T\alpha_0^2 M^2}{16\pi^2(\varepsilon\varepsilon_0)^2 d^6} 
\left\{ 1 + 5\frac{(R_g^{(1)})^2 + (R_g^{(2)})^2}{d^2} \right\}. 
\ee
The vdW interaction between a pair of Gaussian coils is thus attractive (like in the case of stiff polymers), but in contradistinction to the stiff polymer case the interaction is \emph{isotropic}, which we expect because of the sphericity of the coils. 

Now let us consider \emph{identical} Gaussian coils. The polarizabilities are now correlated, and the variance of polarizability fluctuations generates an extra contribution to the sequence average of Eq.~(\ref{eq:Fint_coil}). Consequently, we find 
\be
\langle \langle\!\langle \overline{F_{{\rm int}}} \rangle\!\rangle \rangle_{R_1, R_2} =
-\frac{3 k_{{\rm B}} T(M^2 \alpha_0^2+M g^2)}{16\pi^2(\varepsilon\varepsilon_0)^2 d^6}
\left\{ 1 + 5\frac{(R_g^{(1)})^2 + (R_g^{(2)})^2}{d^2} \right\}. 
\ee
Similar to the case of identical stiff polymers, the effect of the identicality of the Gaussian coils is an overall renormalization of the effective polarizability.     

\subsubsection{Coils near each other}
It is non-trivial to study the interaction of two Gaussian coils whose separation is smaller than the effective radius of gyration, owing to the asphericity of the coils. On the other hand, it is possible to estimate the effective interaction energy of two coils that have a separation distance comparable to their effective radius of gyration, under the condition that $d > 4R_g$. To proceed, let us rewrite the dipole-dipole interaction kernel from Eq.~(\ref{eq:F0_coil}) as follows: 
\ba
&&\int \! \frac{ds}{a}  \! \int  \! \frac{ds'}{a} \frac{1}{|\Rv^{(2)}(s') - \Rv^{(1)}(s)|^6} 
\nonumber\\
&=&
\int \! \frac{ds}{a}  \! \int  \! \frac{ds'}{a} \! \int \! d^3r \! \int \! d^3r' \,  
\frac{\delta(\rv-\Rv^{(1)}(s)) \delta(\rv-\Rv^{(2)}(s'))}{|\rv' - \rv|^6} 
\nonumber\\
&=&
\int \! \frac{ds}{a}  \! \int  \! \frac{ds'}{a} \! \int \! d^3r \! \int \! d^3r' 
\int \! \frac{d^3k}{(2\pi)^3} \! \int \! \frac{d^3k'}{(2\pi)^3} 
\frac{e^{i\kv\cdot(\rv-\Rv^{(1)}(s)) + i\kv'\cdot(\rv'-\Rv^{(2)}(s'))}}{|\rv' - \rv|^6}. 
\label{eq:calc}
\ea
In the second step, we have used the integral representation of the Dirac delta-function. 
For small fluctuations $\delta\Rv^{(i)}(s)$ we can expand the exponent to quadratic order in the fluctuations. Carrying out the thermal average over $\delta\Rv^{(i)}(s)$ and re-exponentiating, Eq.~(\ref{eq:calc}) becomes
\ba
&&M^2\int\frac{ds}{\ell}\int\frac{ds'}{\ell} \!\!
\int \! d^3r \! \int \! d^3r' \!\!
\int \!\! \frac{d^3k}{(2\pi)^3} \! \int \!\! \frac{d^3k'}{(2\pi)^3} \,  
\frac{e^{i\kv\cdot(\rv-\Rv_G^{(1)}) + i\kv'\cdot(\rv'-\Rv_G^{(2)})}}{|\rv' - \rv|^6} 
\nonumber\\
&&\quad\times
\left[ 1 - \frac{1}{2}k_ak_b \! \int \! \frac{ds}{\ell} \langle \delta R_{a}^{(1)}(s) \delta R_{b}^{(1)}(s) \rangle_{R_1}
- \frac{1}{2} k'_a k'_b \! \int \! \frac{ds'}{\ell} \langle \delta R_{a}^{(2)}(s') \delta R_{b}^{(2)}(s') \rangle_{R_2} \right]
\nonumber\\
&=&M^2 \!\! \int \! d^3r \! \int \! d^3r' \!\!
\int \!\! \frac{d^3k}{(2\pi)^3} \! \int \!\! \frac{d^3k'}{(2\pi)^3} \,  
\frac{e^{i\kv\cdot(\rv-\Rv_G^{(1)}) + i\kv'\cdot(\rv'-\Rv_G^{(2)})} 
e^{-\frac{1}{6}k^2(R_g^{(1)})^2 -\frac{1}{6}(k')^2(R_g^{(2)})^2}
}{|\rv' - \rv|^6} 
\nonumber\\
&=&
\frac{27M^2}{8\pi^3 (R_g^{(1)})^3 (R_g^{(2)})^3}
\int \! d^3r \! \int \! d^3r' \, \frac{e^{-\frac{3|\rv-\Rv_G^{(1)}|^2}{2(R_g^{(1)})^2}-\frac{3|\rv'-\Rv_G^{(2)}|^2}{2(R_g^{(2)})^2}}}{|\rv' - \rv|^6}. 
\ea
The interaction is between two polymer segments at $\rv$ and $\rv'$, belonging respectively to the first and second Gaussian coils, and the interactions are weighted by the Gaussian probabilities of finding the corresponding polymer segments away from the centers of mass of their coils. 
The integration is facilitated by making the following change of variables: $\rv = \cv - \frac{1}{2}\lv$, $\rv' = \cv + \frac{1}{2}\lv$. We have 
\ba
&&\int \! \frac{ds}{a}  \! \int  \! \frac{ds'}{a}  
\left\langle \frac{1}{|\Rv^{(2)}(s') - \Rv^{(1)}(s)|^6} \right\rangle_{R_1,R_2}
\nonumber\\
&=&
\frac{27 M^2}{8\pi^3 (R_g^{(1)})^3 (R_g^{(2)})^3}
\int \! d^3c \! \int \! \frac{d^3\lambda}{\lambda^6} \, 
%\nonumber\\
%&&\times
e^{-\frac{3|\cv - \frac{1}{2}\lv-\Rv_G^{(1)}|^2}{2(R_g^{(1)})^2}
-\frac{3|\cv + \frac{1}{2}\lv-\Rv_G^{(2)}|^2}{2(R_g^{(2)})^2}}
\nonumber\\
&=&
\frac{27 M^2}{8\pi^3 (R_g^{(1)})^3 (R_g^{(2)})^3} \left( \frac{2\pi}{3} \right)^{3/2}
\frac{(R_g^{(1)})^3 (R_g^{(2)})^3}{3 \big( (R_g^{(1)})^2 + (R_g^{(2)})^2 \big)^{3/2} } 
%\nonumber\\
%&&\times
\int \! \frac{d^3\lambda}{\lambda^6} \, 
e^{-\frac{|\lv - \dv|^2}{2 R_g^2}}
\nonumber\\
&=&
\frac{M^2}{24\pi^{3/2}R_g^3} \! 
\int \! \frac{d^3\lambda}{\lambda^6} \, 
e^{-\frac{|\lv - \dv|^2}{2 R_g^2}}.
\label{eq:gaussian}
\ea
In the above, we have defined the effective radius of gyration $R_g$ of the two coils, viz., 
\be
R_g^2 \equiv \frac{1}{6}( (R_g^{(1)})^2 + (R_g^{(2)})^2 ).
\label{eq:eff_gyration_r}
\ee 
At room temperature, the radius of gyration is typically of nanometer order for globular proteins; for example, the radius of gyration of lysozyme is approximately $2\,{\rm nm}$ and that of RNA polymerase is approximately $5\,{\rm nm}$.~\cite{nelson} 
In the second step, we have integrated over $\cv$. We see that the integral over $\lambda$ is essentially the dipole-dipole interaction averaged over a Gaussian probability distribution of separation distances that is peaked at a value $\lambda = |\Rv_G^{(2)} - \Rv_G^{(1)}|$, with a variance $\delta\lambda^2 = R_g^2$. Using Eqs.~(\ref{eq:F0_coil}) and (\ref{eq:gaussian}), the interaction energy can be simplified to
\be
\langle\!\langle \overline{F_{{\rm int}}} \rangle\!\rangle = -\frac{k_{{\rm B}} T\alpha_0^2 M^2}{128\pi^{7/2}(\varepsilon\varepsilon_0)^2R_g^3}
\int \! \frac{d^3\lambda}{\lambda^6} \, e^{-\frac{|\lv - \dv|^2}{2R_g^2}}.
\label{eq:Fint_simp}
\ee
\begin{figure}
\begin{center}
		\includegraphics[width=0.5\textwidth]{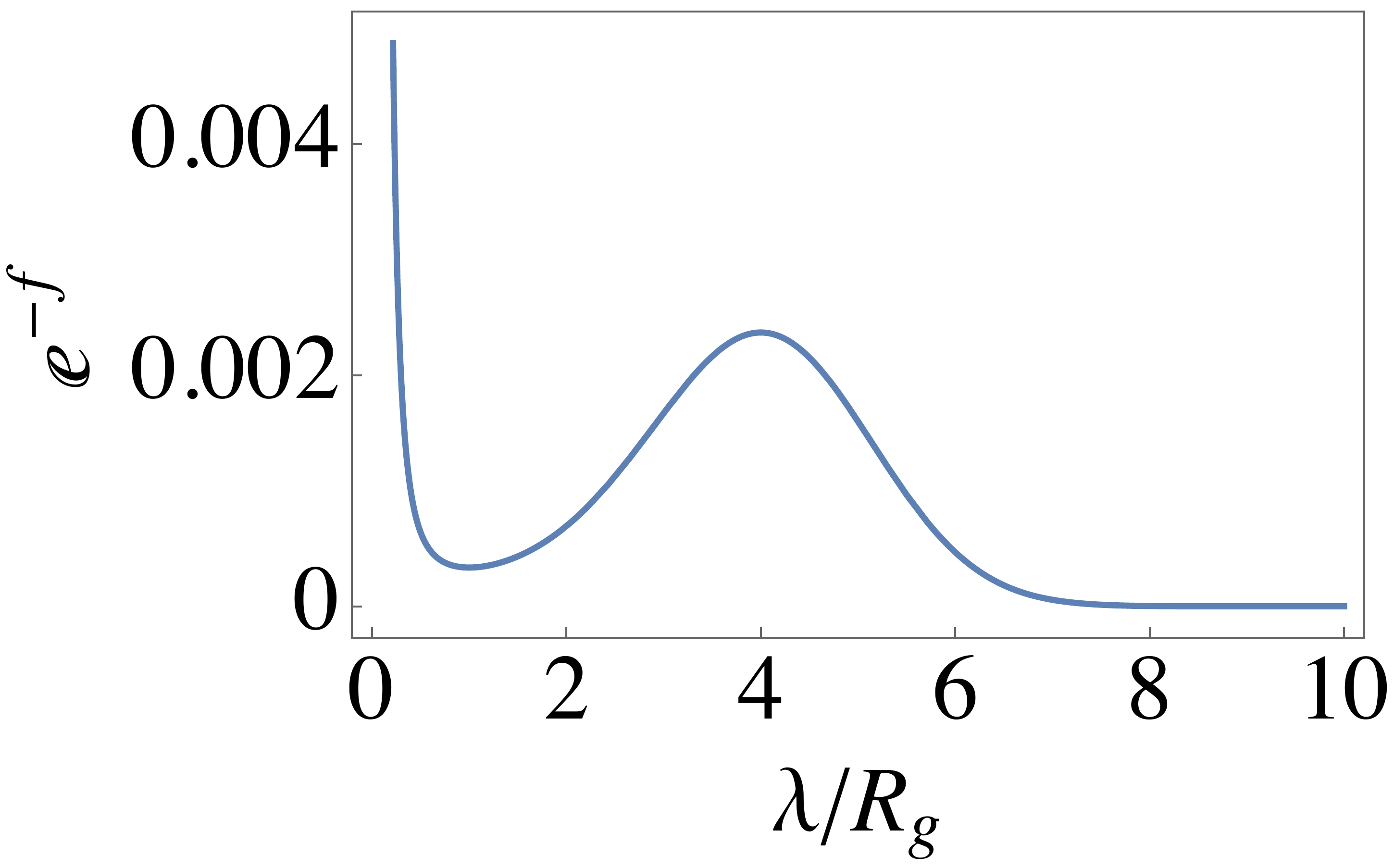}
		\end{center}
	\caption{Behavior of $e^{-f}$ (cf. Eq.~(\ref{eq:f})) as a function of $\lambda/R_g$ for the values $|\Rv_G^{(2)} - \Rv_G^{(1)}|/R_g=5, x=1$. We note the existence of two stationary points: a local maximum (corresponding to $\lambda_+$) and a local minimum (corresponding to $\lambda_-$). If the centers of mass of the two Gaussian coils are farther than $R_g$, the dominant contribution to $e^{-f}$ will come from $\lambda_+$.} 
\label{fig:camel_hump}
\end{figure}

Let us rescale $\lambda \rightarrow R_g \, \lambda$, so that $\lambda$ is dimensionless. We can write
\ba
&&\int \! \frac{d^3\lambda}{\lambda^6} \, e^{-\frac{|\lv - \dv|^2}{2 R_g^2}} \rightarrow  
R_g^{-3} \int \! \frac{d^3\lambda}{\lambda^6} \, e^{-\frac{|\lv - \dv|^2}{2}}
%\nonumber\\
=
2\pi \, R_g^{-3}\! \int_{-1}^{1} \! dx \int_{0}^{\infty} \! d\lambda \, e^{-f},
\label{eq:ef}
\ea
where we have also rescaled $\dv \rightarrow \dv R_g$ (so that $\dv$ is now dimensionless); $x$ is the cosine of the angle between $\lv$ and $\dv$, and $f$ is given by
\be
f = 4 \ln \lambda + \frac{1}{2}(\lambda^2 + d^2 - 2x\lambda \, d).
\label{eq:f}
\ee
Next, we make a saddle point approximation to $\int \! dx \! \int \! d\lambda \, e^{-f(\lambda, x)}$. We approximate the integral by replacing it with the configurations that have $f$ minimized with respect to $x$, i.e., $x=1$. Of these configurations, we approximate the integral over $\lambda$ by the solution to the saddle-point equation, $\frac{\partial f(x=1)}{\partial \lambda}|_{\lambda^\ast} = 0$. 
This yields two solutions, $\lambda_\pm$: 
\be
\lambda_\pm = \frac{d}{2} \left( 1 \pm \sqrt{1- \frac{16}{d^2}} \right).
\ee
We pick the larger one $\lambda_+$ as our saddle point solution, $\lambda_+ \equiv \lambda^\ast$, as it gives a local maximum of $e^{-f}$ (see Fig.~\ref{fig:camel_hump}). The value of $e^{-f}$ diverges at $\lambda=0$, but this is not a problem as we are working in an approximation where the coils should have a minimum separation distance greater than $R_g$. If we restrict the coils to be sufficiently far apart and the effective radius of gyration is sufficiently small, the dominant contribution to the integral will come from $\lambda_+$. 
The solution $\lambda_+$ is real if $d > 4R_g$ (where we have restored dimensional units to $r$). 

In the saddle point approximation, we thus find for the case of coils with distinct sequences
\be
\langle \langle\!\langle \overline{F_{{\rm int}}} \rangle\!\rangle \rangle_{R_1, R_2} 
= 
-\frac{k_{{\rm B}} T\alpha_0^2M^2 C(d,R_g)}{\sqrt{2} \pi^2 (\varepsilon\varepsilon_0)^2 R_g^2} 
%\nonumber\\
%&&\times    
\frac{e^{-\frac{1}{2R_g^2}\big( \frac{1}{2} d^2 - 4 R_g^2 - \frac{1}{2} d^2 \sqrt{1- \frac{16 R_g^2}{d^2}} \big)}}{d^4 \Big( 1 + \sqrt{1- \frac{16 R_g^2}{d^2}} \Big)^4},
\ee
where 
\be
C(d,R_g) \equiv \sqrt{\frac{1 + \sqrt{1-\frac{16R_g^2}{d^2}} - \frac{8R_g^2}{d^2}}{1 + \sqrt{1-\frac{16R_g^2}{d^2}} - \frac{16R_g^2}{d^2}}},
\ee
and for the case of coils with identical sequences
\be
\langle \langle\!\langle \overline{F_{{\rm int}}} \rangle\!\rangle \rangle_{R_1, R_2} 
= 
-\frac{k_{{\rm B}} T(M^2 \alpha_0^2 + M g^2) C(d,R_g)}{\sqrt{2} \pi^2 (\varepsilon\varepsilon_0)^2 R_g^2} 
\frac{e^{-\frac{1}{2R_g^2}\big( \frac{1}{2} d^2 - 4 R_g^2 - \frac{1}{2} d^2 \sqrt{1- \frac{16 R_g^2}{d^2}} \big)}}{d^4 \Big( 1 + \sqrt{1- \frac{16 R_g^2}{d^2}} \Big)^4}.
\ee
The interaction energy of the two identical coils is structurally the same as that of two distinct coils, the only difference being that the prefactor $M^2 \alpha_0^2$ is replaced by $M^2 \alpha_0^2 + M g^2$. A more heterogeneous polarizability sequence thus results in a more attractive vdW interaction between identical coils, but not between distinct coils.

\section{Discussion And Conclusion}
We have considered the pairwise vdW interaction for both stiff polarizable polymers and flexible Gaussian coils, and described their behaviors in the limiting regimes where a pair of polymers are far from each other or near each other. In particular, we see that the vdW interaction between stiff polymers has a marked orientational anisotropy which makes it more attractive for pairs of polymers that are aligned, and moreover we find that in the near-field region, if the stiff polymers have identical sequences, the interaction decays as the inverse fifth power of their separation, which is distinct from and stronger than the inverse fourth power decay of distinct polymers. We have also seen that a more heterogeneous polarizability sequence also results in a more attractive vdW interaction between identical sequences, but not between distinct sequences, for both stiff polymers and flexible Gaussian coils. These characteristic features can aid the mechanism of molecular recognition between biopolymers such as dsDNA molecules in aqueous solvent. 

\section{Acknowledgments}
B.S.L. would like to thank the organizers of the Ninth Alexander Friedmann International Seminar on Gravitation and Cosmology and Third Satellite Symposium on the Casimir Effect for the opportunity to present this paper. B.S.L. and R.P. would like to acknowledge the financial support of the Agency for Research and Development of Slovenia under Grant No. N1-0019.

\end{document}